\documentclass[9pt]{elife}

\usepackage{lipsum} 
\usepackage[version=4]{mhchem}
\usepackage{siunitx}
\usepackage{tikz}
\DeclareSIUnit\Molar{M}
\newcommand{\refstyle}[1]{{\itshape\bfseries\color{eLifeMediumGrey}#1}}
\newcommand{\tikzcircle}[2][red,fill=red]{\tikz[baseline=-0.5ex]\draw[#1,radius=#2] (0,0) circle ;}%
\title{Are theoretical results 'Results'?}

\author[*]{Raymond E. Goldstein}
\affil[]{Department of Applied Mathematics and Theoretical Physics,
Centre for Mathematical Sciences,
University of Cambridge,
Cambridge CB3 0WA, United Kingdom
}

\corr{R.E.Goldstein@damtp.cam.ac.uk}{}

\begin{document}

\maketitle

\begin{abstract}
Yes.
\end{abstract}

\section{Introduction}

The decision letter from the journal was very supportive - it was clear our paper \citep{Kirkegaard} would be published - but one of the referees 
definitely did not like the way we had combined experimental biology and 
physical calculations in our paper: "The data should be described and the inferences drawn, and the modelling relegated to its proper place as quantitative verification of the inferences that can be made directly from the data."

And this was not an isolated case; a referee of another paper had said: "Instead, the authors should let the data speak for itself, and postpone heavier theoretical analysis for later, perhaps in the Discussion."  Many of my colleagues have experienced the same reaction to papers mixing theory and experiment. What were we doing wrong? Why was it not OK, according to these referees, to present the observations and the theory in a back-and-forth dialogue within the 'Results' section?  

While I was bemused by these statements (relegated!), they resonated with my long experience with some in the biology community, namely that they see the significance of theory very differently from the way physicists understand it.  For many biologists, theoretical results are simply not 'Results'.  Indeed, I suspect to many they are seen as a matter of opinion, without any intrinsic significance.  In essence, they don't add anything new.  
Hence the belief in the canonical Results/Discussion dichotomy in which theory (or 'modelling', as it is often called) plays second fiddle, or third.

In contrast,
physicists are brought up to think by means of mathematical 
models: harmonic oscillators, random walks, idealized electrical 
circuits and so on are among the tools in our toolbox, whether we do experiment or theory.  We use them as solvable examples in which a well-defined set of assumptions leads to precise outcomes, and where the dependence of the outcomes on all of the parameters can be interpreted. This approach allows us to estimate what is important and what
is not in any setting. Models also help us to think about problems: "If 
this is the underlying physics, then A should vary with B quadratically...", or
"under these assumptions, the data should collapse like this..." or, when we spot something is not quite right, "here I argue that these claims are in conflict with basic laws of physics" \citep{Meister2016}.

The role of theory is also intimately connected with \textit{predictions}. While I know biologists who would say ``who cares about a prediction in the absence of experiment?”, physicists are brought up to celebrate them - they are the stuff of legend, from 
Dirac’s prediction of antiparticles 
and Einstein’s prediction of the bending of starlight, to the work by 
many that predicted the Higgs particle.  
We view predictions as motivations for experiment and as a means to 
move the discipline forward.  Of course, sometimes they turn out to be wrong, 
but that is often how science works.  
Even if theoretical work does not take the form of a prediction, \textit{per se},
it may still be very useful to design experiments with theory in mind, as emphasized 
by \citet{Bialek}, who has described many historical examples of the role
theory has played in biology, from Rayleigh's work on hearing to Watson and Crick.  

My purpose here is to push back against the view that theory is not a 'Result'.  
I argue for the unabashed inclusion of mathematical formulations and pedagogy within the body of papers published in \textit{eLife} 
and other primarily biological journals.  
By interleaving the experimental and theoretical results it is possible to tell a story, and I firmly believe this makes for much more interesting and readable papers.  It is also faithful to the scientific method, in which one goes back and forth with experiment and hypothesis. 

Readers may be interested to learn that biological information, background and 
results are now routinely included in papers published in physics journals, although this has not always been the case: I vividly recall a situation several decades 
ago when a colleague, a high-energy physicist, saw a preprint about pattern formation in the slime mold \textit{Dictyostelium~discoideum} on my desk and asked: "Why would any physicist study something as ridiculous as that?"  But by now many physicists
do exactly that, and many physics journals are full of discussions of cAMP signaling, 
spiral waves, and chemotaxis \citep{TWC,Rappel,Bodenschatz}.  If we really take 
interdisciplinary research seriously then I assert there has to be a 
prominent place for theory within biology papers, both as Results in papers that combine experiment and theory, and as Results in theory papers.  

This is nothing new.  If you have not already done so, I highly recommend reading the celebrated paper by \citet{Hodgkin} to see experiments and theory interleaved.  Theory is not relegated to the discussion, or worse, to supplementary material, but instead is incorporated into the body of the paper as if it is the most natural thing to do.  And this was in the \textit{Journal of Physiology}. The same structure is found in  the Michaelis-Menten paper, which was published (in German) in a biochemistry journal \citep{MichaelisMenten}; \citep{MichaelisMenten_translation}.  If this was appropriate 
a century ago, why must details of mathematical models now 
be relegated to the back of papers
(see, for example, \citet{Paulick}, \citet{Ferreira}, and \citet{Streichan})?
    
Many readers will appreciate that the issue I am raising about quantitative descriptions of living systems is 
closely associated with the tension that exists between
the stereotypes of the biologist, who wants to incorporate all the complexity of a particular system, and the physicist who seeks generality
and minimalism. As has been emphasized in other recent opinion pieces \citep{Shou, Riveline},
the role of theory in biology has been growing and this development requires
new ways of training scientists on both sides of the physics/biology divide.  
Less attention has been paid to providing concrete examples for the biology
community of how physicists think about understanding data, and this
essay's goal, in part, is to address this lacuna.

Well aware of the risks of trying to speak for an 
entire community, below I take the reader through an
example of how (at least some) physicists might go about describing a well-known phenomenon that shows up everywhere in biology - from the functioning of cellular receptors to bacterial chemotaxis, the propagation of action potentials, and fluorescence recovery after photobleaching (FRAP) - namely, diffusion.  
Employing poetic license, I imagine that we are at a point in 
time when the diffusion equation itself was not known, nor was Fick’s Law, so both the experimental observations and theoretical 
analysis presented below are new and worthy of being described as Results.

I compose two versions of a Results section to indicate various ways of presenting the data and
theory interleaved in a compact presentation that (I hope) is widely understandable by the community.  The first version involves a 'microscopic'  model that is a caricature of the biological system, but contains the essential ingredients to display the behavior observed on the large scale.  The way in which microscopic parameters enter into the macroscopic answer turns out to be general (or, as physicists say, 'universal'), a key 
take-home lesson.  The second version - which is probably more challenging - involves the use of 'dimensional analysis', one of the most powerful methods of analyzing natural phenomena.  Here, relationships between various quantities are deduced by examining the units in which they are measured (mass, length, time, charge, etc.). Introduced long ago, 
particularly in the work of \citet{Maxwell}, this technique can often lead to exact answers to problems, up to the proverbial 'factors of two'.   

\section{A discovery}

\begin{figure}[t]
\centering{\includegraphics[width=100mm]{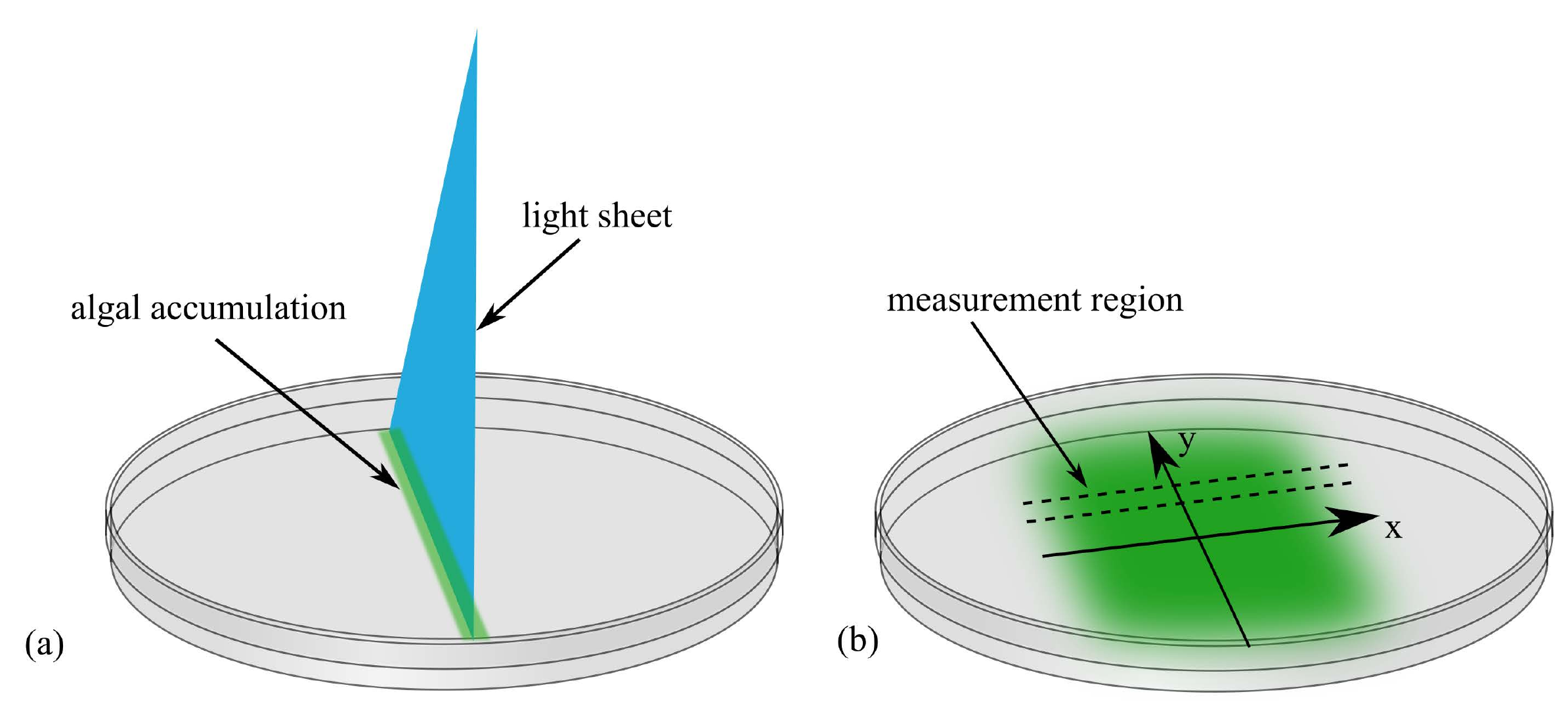}}
\caption{Experimental setup to study diffusion in the green alga \textit{Chlamydomonas}.  (a) A light sheet is used to 
gather the algae, which are swimming in a petri dish, into a narrow strip of cells along the $y$-axis. (b) After the light is turned off, the cells swim randomly and spread out. The concentration profile, $C(x,t)$, is then measured along a thin strip parallel to the $x$-axis; $t$ is time.}
\label{fig:fig1}
\end{figure}

Allow me to introduce our fictitious Professor Lamarr, who has been investigating how the single-cell green alga \textit{Chlamydomonas} move in response to light. She has discovered that if a narrow sheet of light is 
directed into an algal suspension in a petri dish (\FIG{fig1}\refstyle{a}), the algae swim into the beam and form a concentrated line of cells.  When the light is turned off and there is no more phototactic cue, the cells resume a random swimming motion described previously 
\citep{Polin}, in which every $10$ s or so their
roughly linear motion is interrupted by a turn: the angle of this turn falls within a distribution that has a mean of $\sim\!90$ degrees.  These random turns lead the population to spread out over time 
(\FIG{fig1}\refstyle{b}).  See \refstyle{Methods} for experimental details.

\begin{figure}[h]
\centering{\includegraphics[width=120mm]{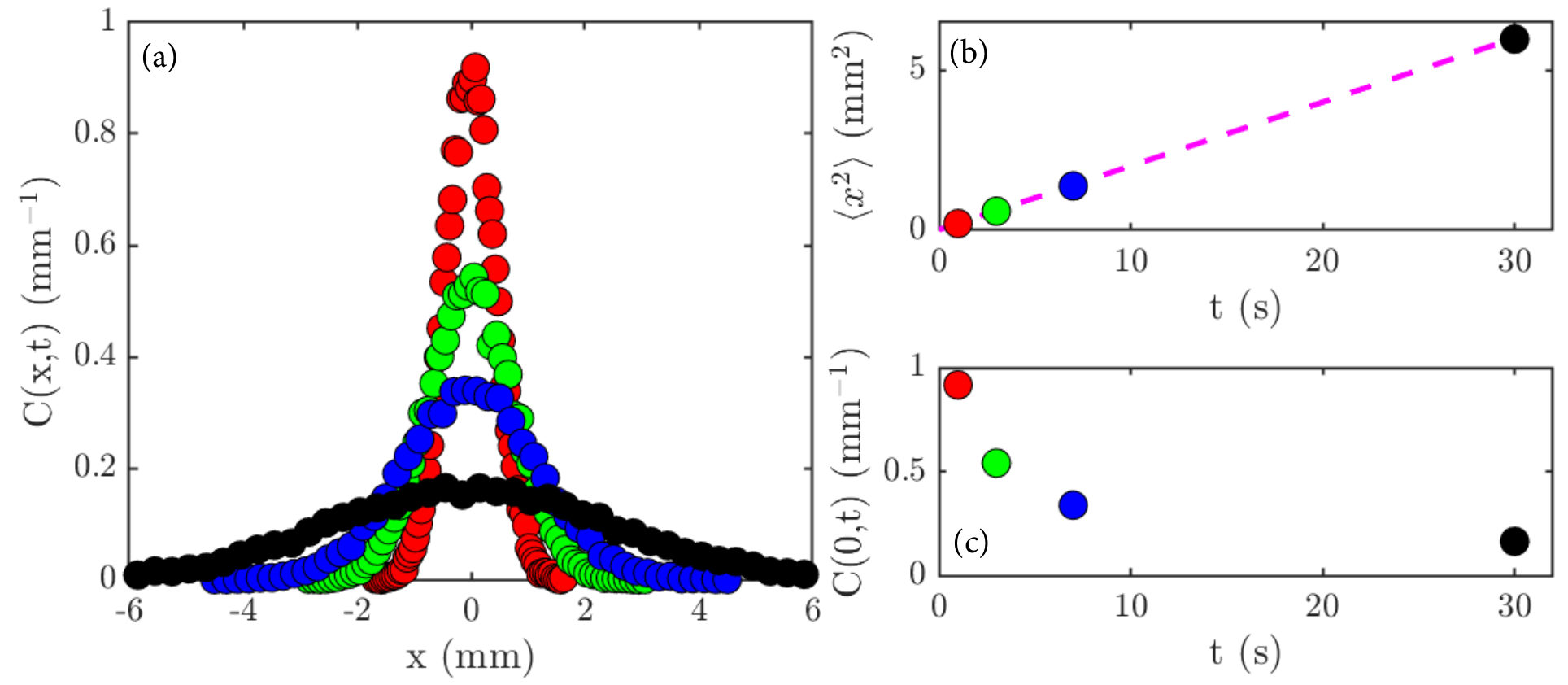}}
\caption{Experimental results on diffusion in a population of the green alga \textit{Chlamydomonas}.  (a) Concentration profiles, $C(x,t)$, normalized to unity, at the following times: 1 second (red), 3 seconds (green), 7 seconds (blue) and 30 seconds (black). (b) The variance, $\langle x^2 \rangle$, of the data shown
in (a) as a function of time; the dashed magenta line is a linear fit to the data. (c) The peak height, $C(0,t)$, of the data shown in (a) as a function of time.}
\label{fig:fig2}
\end{figure}

Lamarr measures the normalized concentration profiles, $C(x,t)$, in a thin strip that is perpendicular to the initial line of cells, obtaining the data shown in \FIG{fig2}\refstyle{a}.  The sharply-peaked profile at early times gradually spreads out until the Petri dish is uniformly filled with cells.  She measured the variance $\langle x^2 \rangle$ of the concentration profile, and found the linear relation 
$\langle x^2 \rangle={\mathscr D}t$, with ${\mathscr D}=0.2$ mm$^2$/s 
(\FIG{fig2}\refstyle{b}). 
Finally, the peak height $C(0,t)$ decays smoothly 
with time (\FIG{fig2}\refstyle{c}). By systematic experimentation, she found that the basic results were insensitive to the precise size of the initial gathering, and that various swimming mutants of \textit{Chlamydomonas} displayed the same behavior, albeit with different values of ${\mathscr D}$.

\subsection{Results v1: Experimental observations explained by a microscopic model}

\begin{figure}[h]
\centering{\includegraphics[width=95mm]{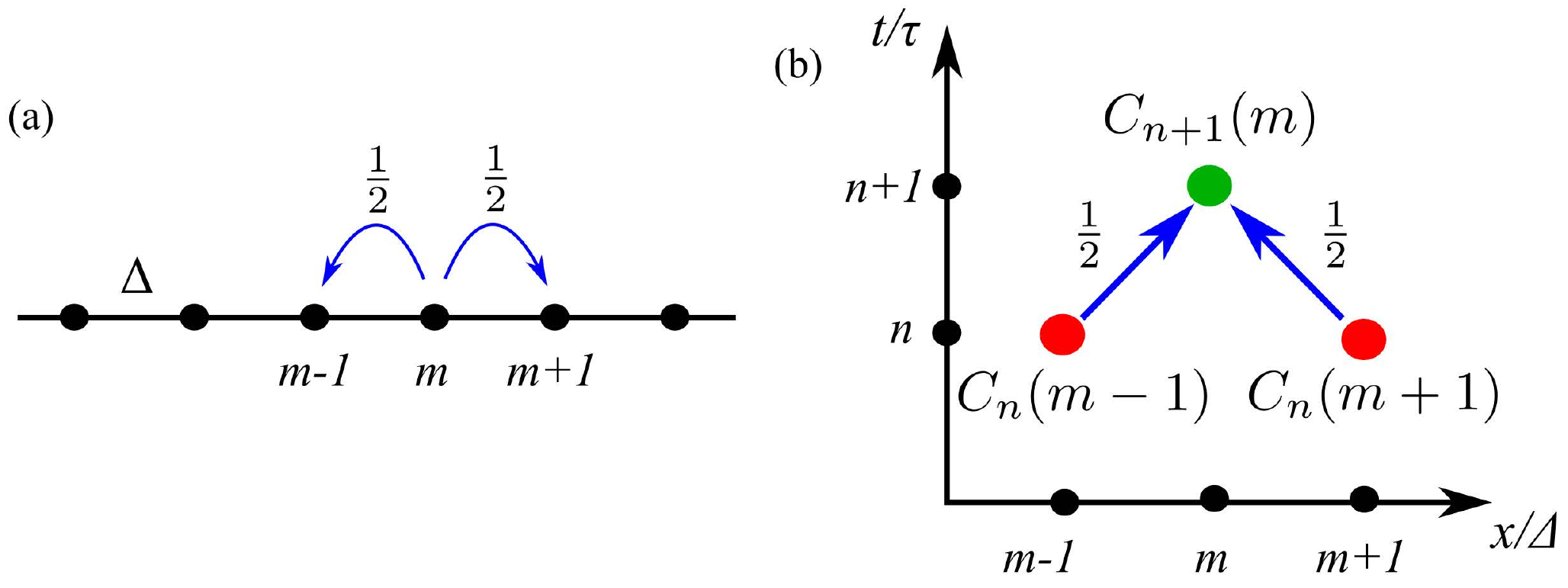}}
\caption{A random walk in one dimension. (a) A cell at site $m$ moves with probability $1/2$ to the left or right. (b) Diagram illustrating
the counting that underlies the evolution equation \eqref{rw1}.}
\label{fig:fig3}
\end{figure}

In this version of Results, we begin with a theoretical model of the random motions of individual cells and deduce from it a population-level description with which to analyze the data. 
In the simplest picture, we assume that cells move only to the left and right along the $x$-axis, and the cells are constrained to sit on 
a discrete set of points, at positions $x_m=m\Delta$, where $m=1,2,3,\ldots$ (\FIG{fig3}\refstyle{a}).  Likewise, we assume time is discrete, so at each time $t_n=n\tau$, $n=1,2,3,\ldots$, 
a cell moves
with probability $1/2$ to the left or right, as indicated by the arrows in \FIG{fig3}\refstyle{a}.

In order to find an evolution equation for the probability $C_n(m)$ of finding a cell at position 
$m\Delta x$
at time $n\Delta t$ we observe (\FIG{fig3}\refstyle{b}) that cells that appear at point
$m$ at time $n+1$ arrived there by moving to the right from point $m-1$ or by moving to the left from point $m+1$ at the previous time step (each with probability $1/2$).  Thus we can deduce that
\begin{equation}
C_{n+1}(m)=\frac{1}{2}C_n(m+1)+\frac{1}{2}C_n(m-1).
\label{rw1}
\end{equation}
We now imagine that the probabilities are varying sufficiently slowly in
space and time that we can use the following Taylor expansions: $C_{n+1}(m)\simeq C_n(m)+\tau (\partial C_n(m)/\partial t) + \cdots$; 
and $C_n(m\pm 1)\simeq C_n(m)\pm 
\Delta (\partial C_n(m)/\partial x)+
(\Delta^2/2)(\partial^2 C_n(m)/\partial x^2) + \cdots$.  Collecting terms, we deduce that the 'continuum limit' for this one-dimensional random walk is
\begin{equation}
\frac{\partial C}{\partial t}= D\frac{\partial^2 C}{\partial x^2}~,\ \ \ \ \ {\rm with}\ \ \ \ \ 
D=\frac{\Delta^2}{2\tau}.
\label{rw3}
\end{equation}
We term this the `diffusion equation', where 
the diffusion constant $D$ has units of length$^2$/time.
Although the above was derived in the context of a model with discrete space and time coordinates, the
crucial point is that we
can more generally interpret $\Delta$ as the typical distance a cell travels between sharp turns, and
$\tau$ as the time between such turns.  If $U$ is the swimming speed between turns, then $\Delta 
\sim U\tau$, so we can write $D=U^2\tau/2$.
From tracking studies of \textit{Chlamydomonas}, we know that $U\sim 0.1$ mm/s, 
and $\tau\sim 10$ s, and therefore $\Delta\sim 1$ mm and $D\sim 0.1$ mm$^2$/s.  

If we rewrite
the diffusion equation \eqref{rw3} as $\partial C/\partial t = 
-(\partial/\partial x) (-D\partial C/\partial x)$ then it can be 
written as
\begin{equation}
\frac{\partial C}{\partial t}=-\frac{\partial J}{\partial x}, \ \ \ \ 
{\rm where} \ \ \ \ J=-D\frac{\partial C}{\partial x},
\label{conservation}
\end{equation}
where we identify the flux $J$ as the number of cells passing through a given point $x$ per unit time.  This relationship implies that cells pass from
regions of high concentration to low at a rate proportional the gradient
of concentration.  
This 'flux form' of
the diffusion equation guarantees that the total number of cells, $N=\int_{-\infty}^{\infty} dx C(x,t)$, remains constant over time, since
\begin{equation}
\frac{dN}{dt}=\int_{-\infty}^{\infty} dx 
\frac{\partial C(x,t)}{\partial t} = 
-\int_{-\infty}^{\infty} dx \frac{\partial J}
{\partial x} = J(-\infty)-J(+\infty).
\label{cons}
\end{equation}
Thus, provided the flux $J$ goes to zero far away from our
point of observation, $N$ is constant.  

\begin{figure}[ht]
\centering{\includegraphics[width=110mm]{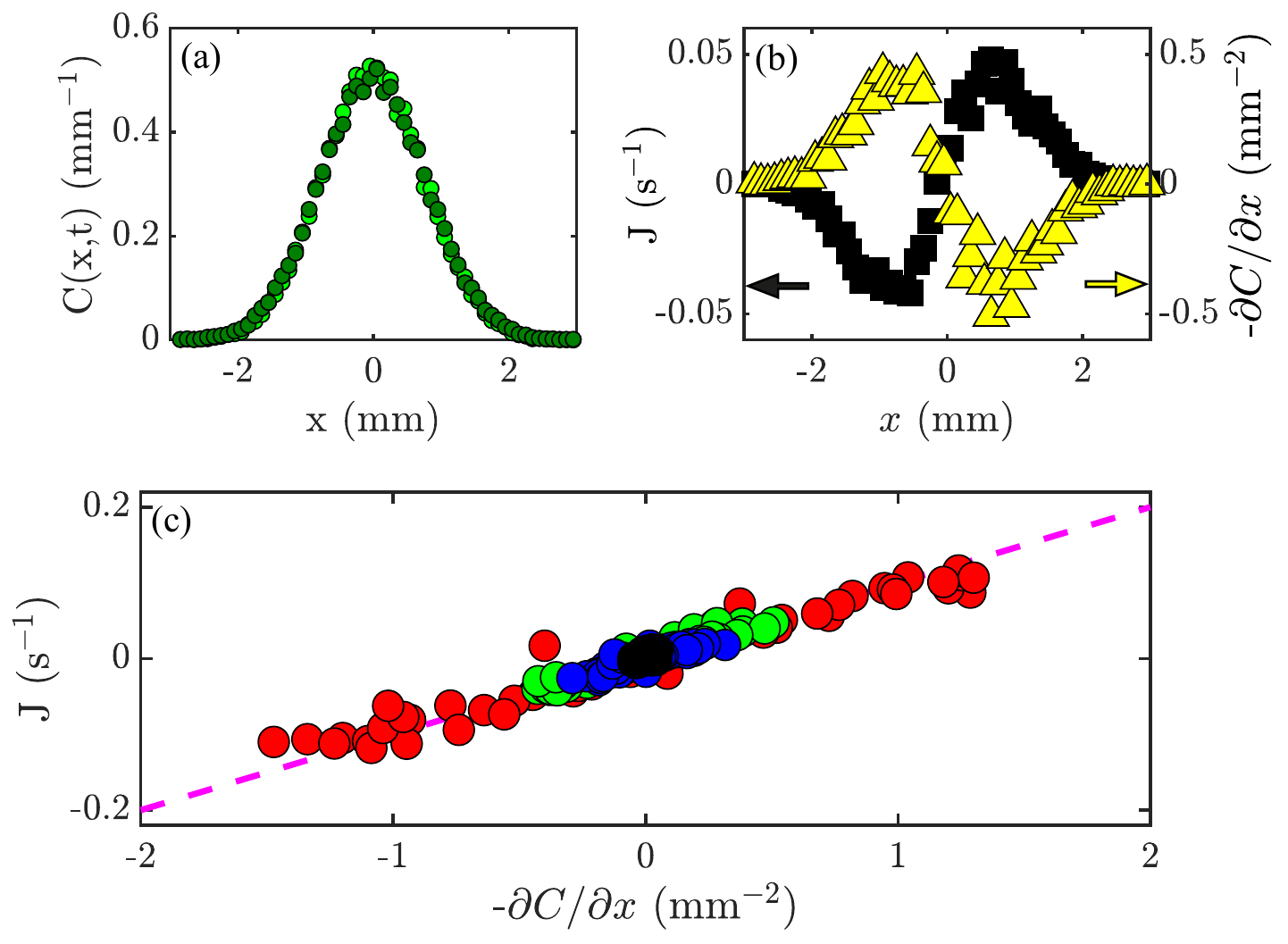}}
\caption{Flux and the diffusion equation.  (a) Concentration profiles, $C(x,t)$, at times $t=3$ \tikzcircle[fill=green]{3pt} and $t=3.2$ s \tikzcircle[fill=black!60!green]{3pt}. (b) The flux of
cells past a given point, $J$ (black; left axis), and the concentration gradient, $\partial C/\partial x$ (yellow; right axis), versus position, $x$. (c) Flux, $J$, versus concentration gradient, $\partial C/\partial x$, for all the values of $x$ and $t$ shown in \FIG{fig2}\refstyle{a}.
The dashed magenta line has a slope $D=0.1$ mm$^2$/s.}
\label{fig:fig4}
\end{figure}

The relationship (Fick's Law) $J = -D \partial C/\partial x$ can be tested experimentally.  Lamarr recorded the distributions of
cells at the times indicated in \FIG{fig2} and
then again $0.2$ s later. As 
shown in \FIG{fig4}\refstyle{a} for one  pair, such
measurements yield the flux, $J$, and concentration gradient, $\partial C/\partial x$ each as functions of $x$ (\FIG{fig4}\refstyle{b}), 
and we see that, apart from the overall scale, they are oppositely signed, as predicted
by \eqref{conservation}. But we can now go one step further and plot $J$ at each point $x$
and time $t$ versus $\partial C/\partial x$ at those same $x$ and $t$ values.  If the theory is correct, then every data set should collapse on to a single
straight line, and indeed this is the case (\FIG{fig4}\refstyle{c}).
According to the theory above, the 
slope of the line in \FIG{fig4}\refstyle{c} is the diffusion constant $D$; we obtain $D=0.1$ mm$^2$/s, which is consistent with the microscopic interpretation in terms of motility.

\subsection{Results v2: Dimensional analysis leads to the diffusion equation}

In this version of the Results section our goal is to infer directly from the data
a differential equation for the time evolution of
the algal concentration $C(x,t)$, which is measured in
organisms per mm, hence units of 1/length. 
The variance $\langle x^2 \rangle$ has, of course, units of length squared, so we can define a characteristic, time-dependent length $\ell(t)=\sqrt{\langle x^2 \rangle}$.
From the fit to the data in \FIG{fig2}\refstyle{b}
we infer that
the width of $C(x,t)$ grows as 
\begin{equation}
\ell(t)\sim \sqrt{{\mathscr D}t}.
\label{variance}
\end{equation}

A very natural question is whether $\ell(t)$ is the only
intrinsic length scale that can be extracted from the data.  As $C(x,t)$
has units of number/length we can, without loss of generality,
write $C(x,t)=\ell(t)^{-1}F(x,t)$ for some unknown function $F$ that
is itself dimensionless. And since $F$ is dimensionless, it must be a function
of a variable that is also dimensionless (similar to the way that $\sin(\theta)$ is a function of $\theta$).
Let us call this dimensionless variable $\xi$.  With $x$ and
$\ell(t)$ to work with, only the ratio is dimensionless, so 
we deduce that $\xi=x/\ell(t)$. 
Thus, we expect
\begin{equation}
C(x,t)=\frac{1}{\ell(t)}F\left(\frac{x}{\ell(t)}\right).
\label{collapse}
\end{equation}

\begin{figure}[h]
\centering{\includegraphics[width=110mm]{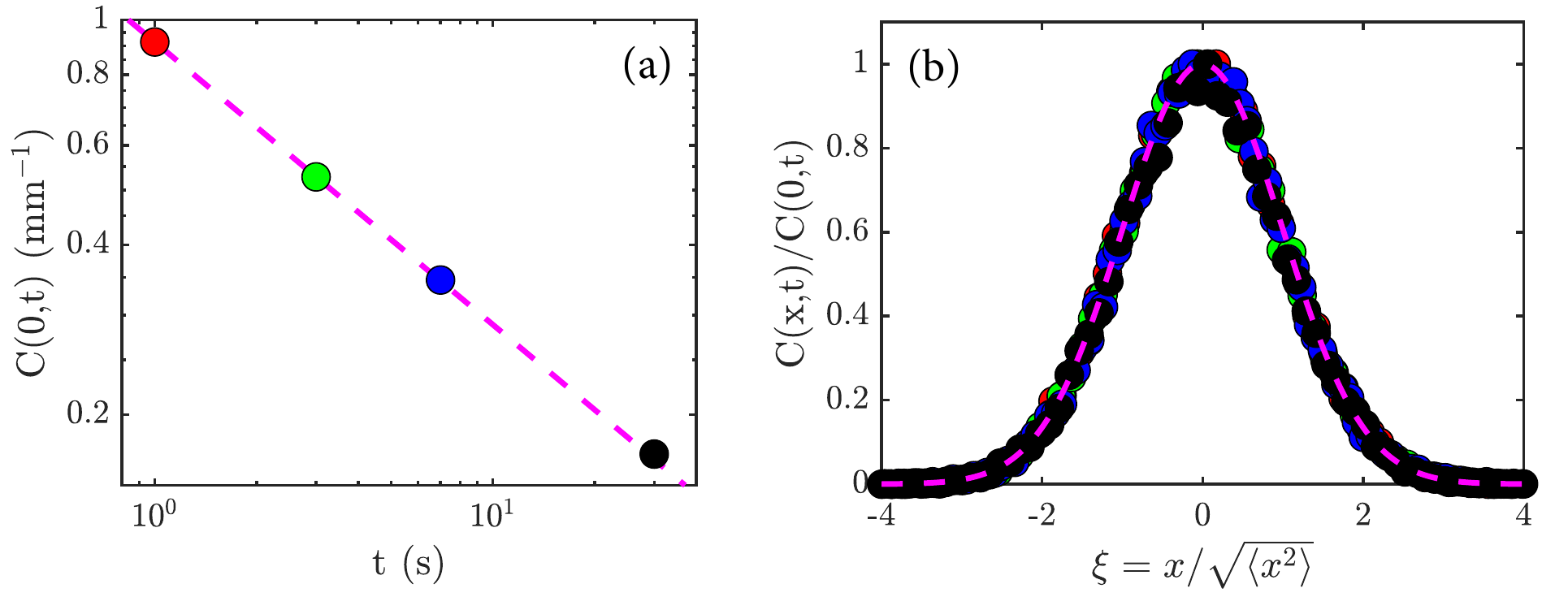}}
\caption{Rescaling the data.  (a) The peak amplitude,$ C(0,t)$, from \FIG{fig2}\refstyle{c} plotted as a function time, $t$, on a log-log scale; the dashed
magenta line has a slope of $-1/2$, which shows that $ C(0,t) \sim t^{-1/2}$.  (b) When the data in  \FIG{fig2}\refstyle{a} are rescaled (see main text) and replotted, they collapse to a universal curve; the dashed magenta curve is the function $\exp(-\xi^2/2)$}
\label{fig:fig5}
\end{figure}

Let us now see if this form is consistent with the data.  First, we note that it guarantees that the total number of cells, $N=\int_{-\infty}^{\infty}\!\! dx~ C(x,t)$, does not change with time because  
\begin{equation}
N=\int_{-\infty}^{\infty}\!\! dx~ C(x,t)=\int_{\infty}^{\infty} dx \frac{1}{\ell(t)}F\left(\frac{x}{\ell(t)}\right)=\int_{-\infty}^{\infty}\! d\xi F(\xi),
\end{equation}
and $\int_{-\infty}^{\infty}\! d\xi F(\xi)$ is a number that does not depend on time (just like $\int_0^\pi d\theta \sin(\theta)$ is a number).  
Given equation \eqref{collapse}, the peak concentration $C(0,t)$ is
just $F(0)/\ell(t)$, where $F(0)$ is again just a number.  With the
scaling in \eqref{variance} we deduce that $C(0,t)\sim 1/\sqrt{t}$.
A replotting of the data in \FIG{fig2}\refstyle{c} on a log-log scale shows
that this is true (\FIG{fig5}\refstyle{a}).  

A significant prediction of the analysis leading to \eqref{collapse} is that the data at different times should collapse when plotted as
$C(x,t)/C(0,t)$ versus
$x/\ell(t)$, for this ratio is just $F(\xi)/F(0)$.  (Dividing $C(x,t)$ by
$C(0,t)$ means that we rescale the heights of the various curves; and dividing $x$ by
$\ell(t)$ means that we allow for expansion of the initial concentration of cells). 
If this holds, then it implies that $\ell(t)$ is the only characteristic length in the system.
A test of this is shown in \FIG{fig5}\refstyle{b}, where we see a good collapse of the data to a universal curve. 

It is natural to seek a differential equation that is consistent with the scaling $x^2\sim t$ and would provide a quantitative prediction of the function $F$. First we consider if inertia is relevant in this system. We know from fluid dynamics that inertia is irrelevant when the Reynolds number $Re=UL/\nu$ is much less than unity: $U$ is the typical speed of a particle, $L$ is the typical length of a particle, and $\nu=\eta/\rho$ is the 
kinematic viscosity (which is defined as $\nu=\eta/\rho$, where $\eta$ is the fluid viscosity and $\rho$ is the fluid density). For \textit{Chlamydomonas} swimming in water ($U \sim 10^{-2}$ cm/s, 
$L\sim 10^{-3}$ cm, and  $\nu=10^{-2}$ cm$^2$/s), we have
$Re\sim 10^{-3}$ and inertia is indeed negligible. 

The differential equation we seek will have derivatives both in time and in
space.  In the absence of inertia, we expect
that the equation for $C(x,t)$ should only
involve first-order derivatives in time (as second derivatives would imply inertia and
accelerations).  With the scaling $x^2\sim t$ we expect two space
derivatives for one time derivative, so a consistent equation would be
\begin{equation}
\frac{\partial C}{\partial t}=D\frac{\partial^2 C}{\partial^2 x},
\label{diffeqn}
\end{equation}
where the parameter $D$ should be proportional to the 
empirical ${\mathscr D}$ obtained from \FIG{fig2}\refstyle{b}.  

To find a solution of \eqref{diffeqn} in the form of \eqref{collapse}, we use $D$ to 
construct a length $l=\sqrt{Dt}$ and find (see \refstyle{Mathematical Details}) the normalized 
distribution
\begin{equation}
C(x,t)=\frac{1}{\sqrt{4\pi Dt}}\exp{\left(-\frac{x^2}{4Dt}\right)}.
\label{green}
\end{equation}
Given this distribution, we compute the variance as
\begin{equation}
\langle x^2 \rangle=\int_{-\infty}^{\infty} x^2 C(x,t)=2Dt.
\label{greenvar}
\end{equation}
Comparing with our empirical observation \eqref{variance}, we deduce the relationship 
${\mathscr D}=2D$ (the promised factor of two!) and therefore that the dimensionless function is 
$F(\xi)=(2\pi)^{-1/2}\exp(-\xi^2/2)$.  The ratio $F(\xi)/F(0)=\exp(-\xi^2/2)$ is shown as
the dashed line in \FIG{fig5}\refstyle{b}, in good agreement
with the data.

Taken together, the experimental observations in \FIG{fig2} and the phenomenological analysis above,
confirmed in \FIG{fig5}, suggest that the diffusion equation in \eqref{diffeqn} provides a sound
description of the spreading of cells that execute random motions.  It indicates
that different organisms, with different diffusion constants, obey the same
fundamental scaling laws, insensitive to the details of the underlying random motions.  Note that at this level of analysis we do not have a microscopic \textit{interpretation} of the diffusion constant in terms
of the fluid viscosity and aspects of cell motility; it is simply a phenomenological parameter that can be used to characterize a given microorganism.  On the other hand, if we knew from
microscopical observations that an organism's motion consists of straight segments
interrupted by random reorientations, as in the case of \textit{Chlamydomonas} 
and indeed \textit{E. coli} \citep{Berg}, then by dimensional analysis (again) we could
deduce $D\sim \Delta^2/\tau\sim U^2\tau$ in terms of the run length $\Delta$, speed $U$,
and time between turns $\tau$.

\section{Discussion}

I have presented two ways of interleaving data and theory in a Results section as a way of indicating how quantitative principles can be used to derive new insight into phenomena.  In one, 
a microscopic model led directly to the diffusion equation, whose structure led to the 'rediscovery'
of Fick's law, which was confirmed from the data.
In the
second, the principles
of dimensional analysis and some phenomenological reasoning led us to
postulate a 'new' diffusion equation as a concise encoding of the experimental observations. 
Each of these approaches used nothing more than basic
algebraic manipulations and elementary differential equations.

Returning to the referees who spoke of inferences drawn directly from the data, I would ask: "What 
language does the
data speak?" The answer would appear to depend on one's background.
The inferences I drew from Lamarr's data were based on experience with understanding continuum and nonequilibrium phenomena, subjects which are less common in the undergraduate physics curriculum than one would hope, and very seldomly found in biology curricula. So, I would indeed advocate a more holistic education for both biologists and physicists \citep{PT}.

It might be argued that the particular example I presented here is unusual, but
in fact these very same considerations (dimensional analysis, scaling collapse of data, etc.) are to be found in many other places in biophysics.  Excellent examples are work on metabolic scaling laws \citep{West} and on stem cell replacement dynamics \citep{Simons}.  

More importantly, I am not trying to emphasize any particular method in the physicist's toolbox, but rather a mindset that is about model-building \textit{and testing} as part of the results presented to the reader.  This mindset is particularly relevant when the theory is formulated first and the experiment is undertaken to test it.  But even when the experiment comes first there may 
be a need to use theory as a sanity check on one's observations \citep{Meister2016}.
This also brings us to the delicate issue of the extent to which research should actually be 'hypothesis driven', as discussed provocatively by \citet{Milner}:  I will leave that Pandora's box closed for the moment. 

Finally, one could argue that the diffusion equation is 'just a model' or 'just a theory' and should, therefore, not be considered as a Result because, unlike the data, it could be shown to be incorrect.  With my experimentalist hat on, I find that argument weak: almost every experiment has potentially confounding aspects, and despite our best efforts to control them, these effects can produce spurious results.  After all, how many hundreds or thousands of papers must have been written about stomach ulcers before \citet{Hpylori} discovered  that \textit{H. pylori} was so often the culprit?  
So, while it is certainly the case that many of the models discussed in biology papers do
not have the status of fundamental laws, I think that it is contrary to the scientific method to view the fact that they may be superseded as a weakness. If theories are
crafted the right way they have utility even if proven wrong, sometimes especially if proven wrong!  

This essay has touched on two tensions - between theory and experiment, and between the cultures of 
physics and biology.  The differences between the cultures have implications not only for
how data is interpreted, but also for what qualifies as "interesting" and
who gets to frame the questions: an enlightening debate on this issue was aired more
than 20 years ago by Adrian Parsegian and Robert Austin \citep{Parsegian,Huebner}.  
For example, it might be argued that biologists may not really be interested in the fact that 
a new equation has been derived that provides an approximate description of a 
given system, and this could be a reason not to publish a theoretical work in a biology journal.  
The example I provide here shows how this need not be an empty exercise, 
but can lead to testable, mechanistic predictions such as the relationship between flux
and concentration gradient (Fick's Law, rediscovered).  One need only consult the seminal work 
of \citet{Turing} on biological pattern formation or of \citet{Hodgkin} on action potentials to 
see the importance of having a mathematical encoding of diffusion to
study its mechanistic implications.  Likewise, a physics-oriented experimental paper, even
one that deals with living organisms, may also not be seen as interesting to biologists because
the questions appear unfamiliar.  For truly interdisciplinary journals, easing this tension
is perhaps the greatest challenge.

\section{Methods}
\subsection{Generating the data} 
Full disclosure - rather than do the experiments, I numerically solved the
Langevin equation $dx/dt=\eta(t)$ for the time evolution of the position $x(t)$ for a 
single alga undergoing random motion,
where $\eta(t)$ is a random variable with zero mean and temporal correlation function
$\langle \eta(t)\eta(t')\rangle=2D\delta(t-t')$.  
In the results described here, I set $D=0.1$ mm$^2$/s, approximately that of \textit{Chlamydomonas} \citep{Polin}. 
The equation was integrated forward a time increment
$\delta t$ from time index $i$ to $i+1$ using the discrete representation $x_{i+1}=x_i+\sqrt{2D\delta t}\eta_i$, where $\eta_i$ is a normally distributed random variable.
The data represent averages over $30,000$ realizations.

\subsection{Mathematical details}
To obtain the normalized concentration profile \eqref{green} we simply substitute the latter into the diffusion equation \eqref{diffeqn}, with $\chi=x/\sqrt{Dt}$.
We obtain
\begin{equation}
\frac{d^2F}{d\chi^2}+\frac{1}{2}\left(F+\chi \frac{dF}{d\chi}\right)=0.
\label{simsoln}
\end{equation}
Integrating \eqref{simsoln} once and imposing the boundary condition that $F\to 0$ as $\chi\to\infty$ we
obtain $dF/d\chi+(1/2)\chi F=0$, which integrates to 
\begin{equation}
F(\xi)=A \exp(-\chi^2/4).
\label{simsoln2}
\end{equation}
Normalizing the associated concentration profile and re-expressing the result in terms of the
original variables yields the result \eqref{green}.

\section{Acknowledgments}

I am grateful to Eric Lauga and Kyriacos Leptos for discussions, to Markus Meister, Philip Nelson, Thomas Powers, Howard Stone, Kirsty Wan, Ned Wingreen, and Francis Woodhouse for reviewing drafts of this essay, This work was supported in part by an Investigator Award from the Wellcome Trust and an Established Career Fellowship from the EPSRC.  Apologies to Betteridge and Hinchliffe for violating their laws of article titles.

\bibliography{resultsresults}

\end{document}